# Variation of the Dependence of the Transient Process Duration on the Initial Conditions in Systems with Discrete Time

A. A. Koronovskiĭ[a,*] and A. E. Hramov[b]

[a] *Saratov State University, Saratov, Russia*
* *e-mail: alkor@cas.ssu.runnet.ru*
[b] *State Scientific Center "College", Saratov, Russia*

**Abstract**—Dependence of the transient process duration on the initial conditions is considered in one- and two-dimensional systems with discrete time, representing a logistic map and the Eno map, respectively.

In the past decades, considerable attention of researchers is devoted to the investigation of nonlinear dynamical systems, both with discrete and continuous time. As a rule, the effort is concentrated on studying established regimes and on determining how one dynamical regime is changed by another in response to variations of the control parameters. At the same time, there are many phenomena related to the transient processes that frequently remain unstudied despite the fact that such processes can provide information about the whole system and about attractors realized in the phase space [1, 2].

Previously [3], we have studied transient processes in a one-dimensional (1D) system with discrete time, representing a logistic map

$$x_{n+1} = f(x_n) = \lambda x_n(1 - x_n). \quad (1)$$

In particular, it was demonstrated that dependence of the transient process duration on the initial conditions $T_\varepsilon(x_0)$ for this logistic map qualitatively changes when the control parameter $\lambda$ varies and one dynamical regime is changed by another. It was also shown that dependence of the transient process duration on the initial conditions obeys certain scaling laws.

In this study, we will consider the mechanisms leading to a qualitative change in the type of dependence of the transient process duration on the initial conditions in the case when the control parameters of a system vary within the framework of the same dynamical regime. The investigation is performed for a logistic map (1), which is a standard object of nonlinear dynamics, and the Eno map [4, 5]

$$\begin{aligned} x_{n+1} &= \lambda x_n(1 - x_n) + b y_n, \\ y_{n+1} &= x_n, \end{aligned} \quad (2)$$

which transforms into a logistic map for $b = 0$. The transient processes in systems (1) and (2) were studied in a simplest regime offered by a stable cycle of period 1. In other words, the control parameters were varied in such a manner that all observations referred to this dynamic regime, in which the behavior of system (2) corresponds to an immobile stable point $(x^0, y^0)$, $x^0 = y^0 = (\lambda + b - 1)/\lambda$, and the behavior of map (1), to an immobile stable point $x^0 = (\lambda - 1)/\lambda$.

Dependence of the transient process duration on the initial conditions, $T_\varepsilon(x_0)$ in the logistic map (1) and $T_\varepsilon(x_0, y_0)$ in the Eno map (2), was determined for the given initial conditions and preset accuracy $\varepsilon$ using the same method as in [3]. First, an attractor realized in the system was determined for a fixed set of the control parameters by $N = 6500$ iterations of an arbitrary initial point, after which it was assumed that the imaging point attained the attractor. Then, the obtained sequence ($\{x_n\}|_{n=0}^N$ and $\{x_n, y_n\}|_{n=0}^N$ for maps (1) and (2), respectively) was analyzed beginning with $n = N - 1$, $N - 2$, … in order to determine a period of the regime (stable point, 2-cycle, 4-cycle, etc.). Finally, by sequentially trying all the possible initial conditions with a certain partition step, an interval of the discrete time necessary for the imaging point to attain the attractor with an accuracy $\varepsilon$ was determined for each initial condition.

The results of this investigation showed that dependence of the transient process duration on the initial conditions $T_\varepsilon(x_0)$ for the logistic map exhibits a qualitative change when the control parameter varies even within the framework of the same dynamicaL regime. Figures 1a and 1b show behavior of the transient process duration $T_\varepsilon(x_0)$ for two values of the control parameter: $\lambda = 1.2$ and $2.9$, respectively. The system behavior in both cases corresponds to the immobile stable point $x_0 = (\lambda - 1)/\lambda$. A change in type of the function $T_\varepsilon(x_0)$ takes place when the control parameter $\lambda$ passes through the value $\lambda_m = 2$, where the cycle of period 1 possesses a maximum stability and the corresponding multiplicator is $\mu = 0$. For $\lambda < \lambda_m$, the function $T_\varepsilon(x_0)$ exhibits two local minima, while for $\lambda > \lambda_m$, the number of such local minima becomes infinite (Fig. 1).

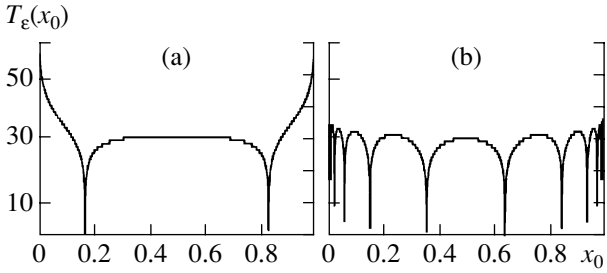

**Fig. 1.** Plots of the transient process duration $T_\varepsilon(x_0)$ as a function of the initial conditions for a cycle of period 1 of the logistic map (1) for two values of the control parameter: $\lambda = 1.2$ (a) and 2.9 (b).

A mechanism of this change in the type of dependence of the transient process duration on the initial conditions is related to the appearance of two sequences of points, $\{x_i^{(0)}\}|_{i=1}^{\infty}$ and $\{x_i^{(1)}\}|_{i=1}^{\infty}$, which are imaged after a finite number of iterations by an immobile stable point $x^0$ such that

$$x^0 = f(x_1^{(0)}) = f(f(x_2^{(0)})) = f(f(x_2^{(1)}))$$
$$= \ldots = f^{(n)}(x_n^{(0)}) = f^{(n)}(x_n^{(1)}). \quad (3)$$

These sequences converge to the points on the boundaries of the basin of attraction for the given attractor, $x_{gr0} = 0$ and $x_{gr1} = 1$, respectively, and each point in the sequence corresponds to a local minimum in the dependence of the transient process duration on the initial conditions (see also [3]). These sequences arise for $\lambda > \lambda_m$ and are related to the irreversibility of map (1).

As noted above, the Eno map (2) transforms for $b = 0$ into the logistic map (1). Therefore, the Eno map with $b = 0$ and $\lambda = 1–3$ must exhibit, within the framework of a stable cycle of period 1, a change in the type of dependence of the transient process duration on the initial conditions $T_\varepsilon(x_0, y_0)$, as is characteristic of the logistic map. In this case, the corresponding dependence $T_\varepsilon(x_0)$ for the logistic map is simply translated along the $y$ axis so that the local minima in the dependence of the transient process duration on the initial conditions pass by the lines parallel to the $y$ axis on the plane of initial conditions. This phenomenon is also observed in the Eno map for the values of the control parameter $b$ close (but not equal) to zero (Fig. 2). The appearance of a

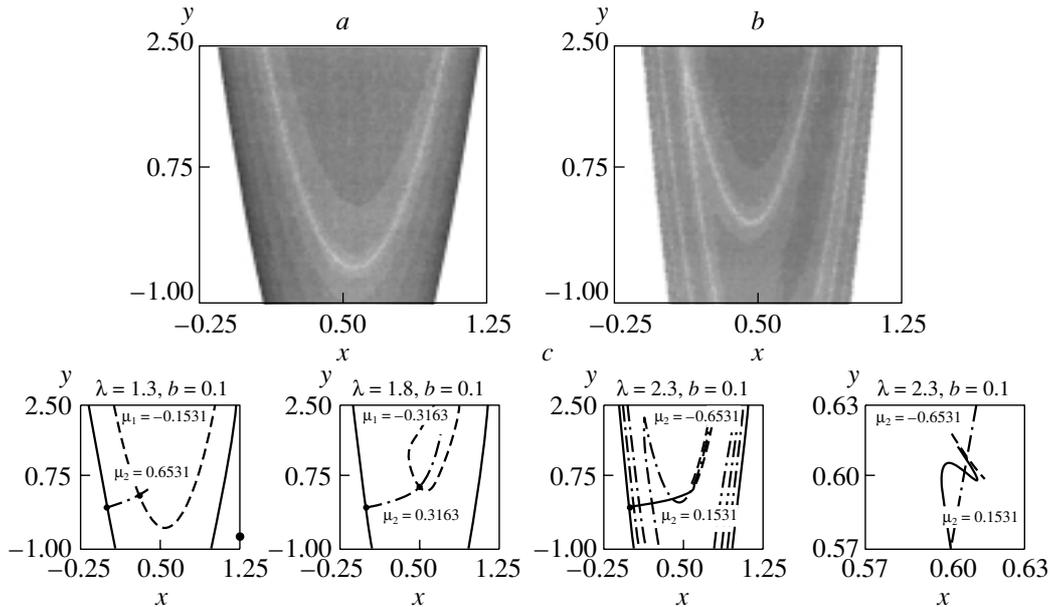

**Fig. 2.** (a, b) Projection of the surface $T_\varepsilon(x_0, y_0)$, representing the transient process duration as a function of the initial conditions, on the plane of possible states $(x, y)$ for the Eno map with $b = 0.1$ and $\lambda = 1.3$ and 2.3, respectively. Variations in the transient process duration are reflected by the gray color gradations: from white (corresponding to $T = 0$) to black (corresponding to $T = 50$ units of the discrete time). A light-gray shade outside the basin of attraction of the attractor $(x^0, y^0)$ corresponds to the initial conditions with an attractor at the infinity.

(c) Schematic diagrams showing stable manifolds of the attractor point $(x^0, y^0)$ and the stable and unstable manifolds of the instable point (0, 0) for a fixed value of the control parameter ($b = 0.1$) and different values of the parameter $\lambda$. For the point $(x^0, y^0)$, stable manifolds are depicted by dashed curve for a negative multiplicator $\mu_1$ and by dot-dash curve for a positive multiplicator $\mu_2$ (the values of $\mu_1$ and $\mu_2$ are indicated at the curves). Solid curves show the manifolds for the unstable point (0, 0). The last diagram shows a magnified fragment of the previous diagram, refining behavior of the unstable manifold of the point (0, 0) in the vicinity of the attractor $(x^0, y^0)$.

nonzero term $by_n$ in Eq. (2) results in that the lines corresponding to minima of the $T_\varepsilon(x_0, y_0)$ function are no longer parallel.

At the same time, it is not possible to explain the mechanism of complication of the dependence of the transient process duration on the initial conditions $T_\varepsilon(x_0, y_0)$ in the same way as it was done above for the logistic map. This is related, first, to the fact that the Eno map with $b \neq 0$ is reversible and, hence, there are no points imaged by an immobile point $(x^0, y^0)$ after a finite number of iterations.[1] However, it was just the existence of such points that accounted for the appearance of the infinite number of local minima of the $T_\varepsilon(x_0)$ function for the logistic map.

Second, in the Eno map with $b \neq 0$, the multiplicators $\mu_{1,2}$ of an immobile point $(x_0, y_0)$ do not take zero values and, accordingly, the Eno map (with $b \neq 0$) possess no cycles of maximum stability. However, the bifurcation in dependence of the transient process duration $T_\varepsilon(x_0)$ on the initial conditions for the logistic map was related just to the cycle of maximum stability. Therefore, the Eno map leads to the problem of finding a bifurcation line on the plane of control parameters $(\lambda, b)$, the crossing of which results in complication of the shape of the $T_\varepsilon(x_0, y_0)$ function.

Let us compare of the projection of the $T_\varepsilon(x_0, y_0)$ surface onto the plane of initial conditions $(x_0, y_0)$ to the arrangement of stable manifolds of a stable immobile point $(x^0, y^0)$. Such manifolds play an important role, for example, in the formation of boundaries of the basins of attraction for attractors [6, 7]. This comparative analysis shows that the points of initial conditions for which the transient process duration is minimum coincide with the points of one of the manifolds. A stable immobile point $(x^0, y^0)$ is characterized by the multiplicators $\mu_1 = (2 - 2b - \lambda - \sqrt{4b + (-2 + 2b + \lambda)^2})/2)$ and $\mu_2 = (2 - 2b - \lambda + \sqrt{4b + (-2 + 2b + \lambda)^2})/2)$, as well as by the corresponding eigenvectors $\mathbf{e}_1 = (2 - 2b - \lambda - \sqrt{4b + (-2 + 2b + \lambda)^2})/2$ and $\mathbf{e}_2 = (2 - 2b - \lambda + \sqrt{4b + (-2 + 2b + \lambda)^2})/2$.

Minimum values of the function $T_\varepsilon(x_0, y_0)$ are attained at that manifold of the point $(x^0, y^0)$ which corresponds to the minimum value of multiplicator. This manifold provides for the most rapid approach of the imaging point to the attractor. Variation of the control parameter $\lambda$ from 0.9 to 0.27 at a fixed value of the control parameter $b = 0.1$ (whereby behavior of the system corresponds to a stable cycle of period 1) leads to a gradual decrease of the value of the positive multiplicator $\mu_2$ and an increase in the absolute value of the negative multiplicator $\mu_1$. For $b = 0.1$ and $\lambda < \lambda_{eq} = 1.8$, the system is characterized by $|\mu_1| < |\mu_2|$ and, accordingly, the minima in the dependence of the transient process duration on the initial conditions (represented by light bands in Fig. 2a) correspond to a manifold characterized by the negative multiplicator $\mu_1$. At $\lambda = \lambda_{eq} = 1.8$, the multiplicators are equal ($|\mu_1| = |\mu_2|$) and the two manifolds become "equivalent". For $\lambda > \lambda_{eq} = 1.8$, we obtain $|\mu_2| < |\mu_1|$ and the manifold change in their roles: the light bands of minimum values of the transient process durations correspond to the manifold characterized by a positive multiplicator $\mu_2$ (Fig. 2b).

Thus, the control parameters $b = 0.1$, $\lambda = 1.8$ correspond to a situation when the stable manifolds of an immobile stable point $(x^0, y^0)$ "change their roles." However, the same values of the control parameters correspond to another important event, involving unstable manifold of the unstable point (0, 0). Note that a stable manifold of this point, characterized by the multiplicator $\mu_1^{unst} = (\lambda - \sqrt{4b + \lambda^2})/2$ and the eigenvector $\mathbf{e}_1 = (\lambda - \sqrt{4b + \lambda^2})/2$, forms a boundary of the basin of attraction of the attractor $(x^0, y^0)$. The unstable manifold of the point (0, 0) is characterized by the multiplicator $\mu_2^{unst} = (\lambda + \sqrt{4b + \lambda^2})/2$ and the eigenvector $\mathbf{e}_2 = (\lambda + \sqrt{4b + \lambda^2})/2$. Since any point occurring in a small vicinity of the unstable point (0, 0) but belonging to the basin of attraction of the attractor $(x^0, y^0)$ must be attracted to the latter with time, the unstable manifold of the unstable point (0, 0) must terminate at the point $(x^0, y^0)$ (or at least pass through this point).

As noted above, for $\lambda < \lambda_{eq} = 1.8$, a dominating role belongs to a manifold corresponding to the negative multiplicator $\mu_1$. It is this manifold, on which the most rapid approach to the attractor takes place. Therefore, the unstable manifold of the point (0, 0) tends to the attractor along the second manifold of the point $(x_0, y_0)$ characterized by the multiplicator $\mu_2$ (Fig. 2c). When the control parameter $\lambda_2$ exceeds the value $\lambda_{eq}$, behavior of the unstable manifold of the point (0, 0) in the vicinity of the point $(x^0, y^0)$ is determined by a manifold with the negative multiplicator $\mu_1$. As a result, the unstable manifold of the point (0, 0) crosses for the infinite number of times the stable manifold of $(x^0, y^0)$ characterized by the positive multiplicator $\mu_2$. Therefore, the stable manifold of the point $(x^0, y^0)$ also crosses for the infinite number of times the unstable manifold of (0, 0) in the vicinity of this point. This leads to the appearance of the infinite number of "depressions" on the $T_\varepsilon(x_0, y_0)$ surface.

Thus, under the condition $\mu_1 = -\mu_2$ (both $\mu_1$ and $\mu_2$ are real), the Eno map exhibits a bifurcation of manifold splitting. This leads to a qualitative change in dependence of the transient process duration on the initial conditions.

---
[1] Naturally, except the point $(x^0, y^0)$ itself.

**Acknowledgments.** This study was supported by the Russian Foundation for Basic Research (project nos. 01-02-17392 and 00-15- 96673) and by the Scientific-Education Center "Nonlinear Dynamics and Biophysics" at the Saratov State University (Grant REC-006 from the US Civilian Research and Development Foundation for Independent States of the Former Soviet Union).